\documentclass{ifacconf}
\usepackage{amsmath,amssymb,amsfonts} 
\usepackage{mathtools} 
\usepackage{xfrac} 
\usepackage{siunitx} 
\newcommand{\slfrac}[2]{\left.#1\middle/#2\right.}
\usepackage{enumerate} 
\usepackage[shortlabels]{enumitem}
\usepackage{graphicx} 
\graphicspath{ {imgs/} } 
\usepackage{textcomp} 
\usepackage{physics} 
\usepackage{xcolor} 
\usepackage{subfiles} 

\usepackage[algo2e,ruled]{algorithm2e}
\SetKwComment{Comment}{/* }{ */} 

\usepackage{lmodern} 
\usepackage{anyfontsize} 

\usepackage{natbib}
\usepackage{url} 

\counterwithin*{section}{part}
\usepackage{cleveref}

\usepackage{microtype}

\usepackage{amsmath}
\usepackage{xifthen} 
\usepackage{xstring} 
\usepackage{xparse} 

\let\bfvar\undefined\NewDocumentCommand{\bfvar}{m O{} O{} O{} }{
	\def\basetmp{\boldsymbol{#1}}%
	\ifthenelse{\isempty{#4}}{\def\base{\basetmp}}{\def\base{#4{\basetmp}}}%
	\ensuremath{\base{}\ifthenelse{\isempty{#3}}{}{^{#3}}\ifthenelse{\isempty{#2}}{}{_{#2}}}%
}

\let\originalleft\left
\let\originalright\right
\renewcommand{\left}{\mathopen{}\mathclose\bgroup\originalleft}
\renewcommand{\right}{\aftergroup\egroup\originalright}

\DeclareDocumentCommand{\abs}{s m}{
	\IfBooleanTF{#1}
	{\ensuremath{|#2|}} 
	{\ensuremath{\left|#2\right|}} 
}
\DeclareDocumentCommand{\norm}{s m}{
	\IfBooleanTF{#1}
	{\ensuremath{\|#2\|}} 
	{\ensuremath{\left\|#2\right\|}} 
}
\DeclareDocumentCommand{\para}{s m}{
	\IfBooleanTF{#1}
	{\ensuremath{(#2)}} 
	{\ensuremath{\left(#2\right)}} 
}
\DeclareDocumentCommand{\curl}{s m}{
	\IfBooleanTF{#1}
	{\ensuremath{\{#2\}}} 
	{\ensuremath{\left\{#2\right\}}} 
}
\DeclareDocumentCommand{\bracc}{s m}{
	\IfBooleanTF{#1}
	{\ensuremath{[#2]}} 
	{\ensuremath{\left[#2\right]}} 
}

\DeclareDocumentCommand{\det}{s m}{
	\IfBooleanTF{#1}
	{\ensuremath{\textsf{det}\para*{#2}}} 
	{\ensuremath{\textsf{det}\para{#2}}} 
}
\DeclareDocumentCommand{\diag}{s m}{
	\IfBooleanTF{#1}
	{\ensuremath{\textsf{diag}\para*{#2}}} 
	{\ensuremath{\textsf{diag}\para{#2}}} 
}
\DeclareDocumentCommand{\trace}{s m}{
	\IfBooleanTF{#1}
	{\ensuremath{\textsf{tr}\para*{#2}}} 
	{\ensuremath{\textsf{tr}\para{#2}}} 
}
\DeclareDocumentCommand{\exp}{s m}{
	\IfBooleanTF{#1}
	{\ensuremath{\textrm{exp}\para*{#2}}} 
	{\ensuremath{\textrm{exp}\para{#2}}} 
}
\DeclareDocumentCommand{\vectrafo}{s m}{
	\IfBooleanTF{#1}
	{\ensuremath{\textsf{vec}\para*{#2}}} 
	{\ensuremath{\textsf{vec}\para{#2}}} 
}
\DeclareDocumentCommand{\skew}{s m}{
	\IfBooleanTF{#1}
	{\ensuremath{\textsf{sk}\para*{#2}}} 
	{\ensuremath{\textsf{sk}\para{#2}}} 
}
\DeclareDocumentCommand{\dotprod}{s m m}{
	\IfBooleanTF{#1}
	{\ensuremath{\langle#2,#3\rangle}} 
	{\ensuremath{\left\langle#2,#3\right\rangle}} 
}

\let\Lip\undefined\NewDocumentCommand{\Lip}{ O{} }{
	\ensuremath{L\ifthenelse{\isempty{#1}}{}{_{#1}}}%
}

\let\Rspace\undefined\newcommand{\Rspace}[1][]{\ensuremath{\mathbb{R}\ifthenelse{\isempty{#1}}{}{^{#1}}}} 
\let\SEthree\undefined\newcommand{\SEthree}{\ensuremath{SE(3)}} 
\let\sethree\undefined\newcommand{\sethree}{\ensuremath{se(3)}} 
\let\SOthree\undefined\newcommand{\SOthree}{\ensuremath{SO(3)}} 

\let\Dset\undefined\newcommand{\Dset}{\mathcal{D}} 
\let\Xset\undefined\newcommand{\Xset}{\mathcal{X}} 
\let\bodyvelclass\undefined\newcommand{\bodyvelclass}{\ensuremath{\mathcal{V}^b}}
\let\Bset\undefined\newcommand{\Bset}{\mathbb{B}}

\let\eye\undefined\NewDocumentCommand{\eye}{ O{} }{\bfvar{I}[#1]}
\let\jacobian\undefined\newcommand{\jacobian}{\ensuremath{\boldsymbol{J}}}
\let\adjoint\undefined\NewDocumentCommand{\adjoint}{s O{} O{} }{\bfvar{\mathrm{Ad}}[\ifthenelse{\isempty{#2}}{}{\IfBooleanTF{#1}{\para*{#2}}{\para{#2}}}][][#3]}
\let\N\undefined\newcommand{\N}{\ensuremath{\boldsymbol{\mathrm{N}}}}

\let\frame\undefined\NewDocumentCommand{\frame}{ O{} }{\bfvar{\mathrm{\Sigma}}[#1][]}

\let\g\undefined\NewDocumentCommand{\g}{ O{} O{} }{\bfvar{g}[#1][][#2]}
\let\gcheck\undefined\NewDocumentCommand{\gcheck}{ O{} O{} }{\bfvar{\mathfrak{g}}[#1][][#2]}
\let\p\undefined\NewDocumentCommand{\p}{ O{} O{} }{\bfvar{p}[#1][][#2]}
\let\eulax\undefined\NewDocumentCommand{\eulax}{ O{} O{} }{\bfvar{\xi}[#1][][#2]}
\let\axang\undefined\NewDocumentCommand{\axang}{ O{} O{} }{\eulax[][#2]\theta\ifthenelse{\isempty{#1}}{}{_{#1}}}
\let\rot\undefined\NewDocumentCommand{\rot}{ O{} }{
	\StrLeft{#1}{1}[\stringsign]%
	\StrGobbleLeft{#1}{1}[\stringwithoutsign]%
	\ifthenelse{ \equal{\stringsign}{-} }%
	{\ensuremath{e^{-\axang[\stringwithoutsign][\hat]}}}%
	{\ensuremath{e^{\axang[#1][\hat]}}}%
}

\let\bodyvel\undefined\NewDocumentCommand{\bodyvel}{ O{} O{} }{\bfvar{V}[#1][b][#2]}
\let\bodyvelH\undefined\NewDocumentCommand{\bodyvelH}{ O{} }{\bodyvel[#1][\hat]}
\let\transvel\undefined\NewDocumentCommand{\transvel}{ O{} }{\bfvar{v}[#1][b]}
\let\angvel\undefined\NewDocumentCommand{\angvel}{ O{} O{} }{\bfvar{\omega}[#1][b][#2]}
\let\angvelskew\undefined\NewDocumentCommand{\angvelskew}{ O{} }{\angvel[#1][\hat]}

\let\gbar\undefined\NewDocumentCommand{\gbar}{ O{} O{} }{
	\def\basetmp{\bar{\boldsymbol{g}}}%
	\ifthenelse{\isempty{#2}}{\def\base{\basetmp}}{\def\base{#2{\basetmp}}}%
	\ensuremath{\base{}\ifthenelse{\isempty{#1}}{}{_{#1}}}%
}
\let\gcheckbar\undefined\NewDocumentCommand{\gcheckbar}{ O{} O{} }{
	\def\basetmp{\bar{\boldsymbol{\mathfrak{g}}}}%
	\ifthenelse{\isempty{#2}}{\def\base{\basetmp}}{\def\base{#2{\basetmp}}}%
	\ensuremath{\base{}\ifthenelse{\isempty{#1}}{}{_{#1}}}%
}
\let\pbar\undefined\NewDocumentCommand{\pbar}{ O{} }{\p[#1][\bar]}
\let\bodyvelbar\undefined\NewDocumentCommand{\bodyvelbar}{ O{} }{\bodyvel[#1][\bar]}

\let\u\undefined\NewDocumentCommand{\u}{ O{} }{\ensuremath{\boldsymbol{u}\ifthenelse{\isempty{#1}}{}{_{#1}}}}
\let\e\undefined\NewDocumentCommand{\e}{ O{} }{\ensuremath{\boldsymbol{e}\ifthenelse{\isempty{#1}}{}{_{#1}}}}
\let\nuu\undefined\NewDocumentCommand{\nuu}{ O{} }{\ensuremath{\boldsymbol{\nu}\ifthenelse{\isempty{#1}}{}{_{#1}}}}
\let\vismea\undefined\NewDocumentCommand{\vismea}{ O{} O{} }{\bfvar{f}[\ifthenelse{\isempty{#1}}{}{#1}][][#2]}

\let\K\undefined\NewDocumentCommand{\K}{ O{} }{\ensuremath{\boldsymbol{K}\ifthenelse{\isempty{#1}}{}{_{#1}}}}

\let\x\undefined\newcommand{\x}{\ensuremath{\boldsymbol{x}}}
\let\y\undefined\newcommand{\y}{\ensuremath{\boldsymbol{y}}}
\let\eps\undefined\newcommand{\eps}{\ensuremath{\boldsymbol{\epsilon}}}
\let\Xdata\undefined\newcommand{\Xdata}{\boldsymbol{X}} 
\let\Ydata\undefined\newcommand{\Ydata}{\boldsymbol{Y}} 

\let\Nspace\undefined\NewDocumentCommand{\Nspace}{s m m}{ 
	\IfBooleanTF{#1}
	{\ensuremath{\mathcal{N}\para*{#2,\, #3}}} 
	{\ensuremath{\mathcal{N}\para{#2,\, #3}}} 
}
\let\Probspace\undefined\NewDocumentCommand{\Probspace}{s m}{\ensuremath{\mathsf{Pr}\IfBooleanTF{#1}{\curl*{#2}}{\curl{#2}}}} 

\let\betaGP\undefined\newcommand{\betaGP}{\boldsymbol{\beta}} 
\let\maxinfoGP\undefined\newcommand{\maxinfoGP}{\boldsymbol{\zeta}} 

\let\kernel\undefined\newcommand{\kernel}{\ensuremath{\mathrm{k}}} 
\let\Gram\undefined\newcommand{\Gram}{\ensuremath{\boldsymbol{\mathrm{K}}}} 
\let\extendedcov\undefined\newcommand{\extendedcov}{\ensuremath{\boldsymbol{\mathrm{k}}}} 
\let\meanGP\undefined\newcommand{\meanGP}{\ensuremath{\boldsymbol{\mu}}} 
\let\varGP\undefined\newcommand{\varGP}{\ensuremath{\boldsymbol{\Sigma}}} 

\DeclareMathOperator*{\argmin}{argmin}
\DeclareMathOperator{\atantwo}{arctan2}

\newtheorem{assumption}{Assumption}
\newtheorem{remark}{Remark}
\newtheorem{theorem}{Theorem}
\newtheorem{lemma}{Lemma}
\newtheorem{corollary}{Corollary}[theorem]

\newenvironment{proof}{\noindent\hspace{0em}{\textit{Proof:}} }{\hspace*{\fill}~$\blacksquare$\par\endtrivlist\unskip}

\crefname{equation}{}{}

\crefname{enumi}{}{}
\Crefname{enumi}{}{}

\crefname{assumption}{ass.}{ass.}
\Crefname{assumption}{Assumption}{Assumptions}

\crefname{algorithm}{alg.}{algs.}
\Crefname{algorithm}{Algorithm}{Algorithms}

\Crefname{figure}{Fig.}{Figs.}

\Crefname{remark}{Remark}{Remarks}

\crefname{corollary}{cor.}{cor.}
\Crefname{corollary}{Corollary}{Corollaries}

\let\LambdaGP\undefined\newcommand{\LambdaGP}{\ensuremath{\boldsymbol{\Lambda}}}
\let\GPmodel\undefined\newcommand{\GPmodel}[1][]{\ensuremath{\mathcal{GP}\ifthenelse{\isempty{#1}}{}{_{#1}}}}
\let\modelerror\undefined\newcommand{\modelerror}{\ensuremath{\Delta_{\bfvar{V}[\psi]}}} 
\let\maxmodelerror\undefined\newcommand{\maxmodelerror}{\ensuremath{\bar{\Delta}{}_{\bfvar{V}}}} 
\let\alphaswitch\undefined\newcommand{\alphaswitch}{\ensuremath{\boldsymbol{\alpha}}}

\let\MATLAB\undefined\newcommand{\MATLAB}{{\fontfamily{ptm}\selectfont{MATLAB}}}

\begin{document}
	
\begin{frontmatter}
	
	\title{Visual Pursuit Control based on Gaussian Processes with Switched Motion Trajectories\thanksref{footnoteinfo}} 
	
	\thanks[footnoteinfo]{This work was supported by JSPS KAKENHI Grant Number 20K14761.}
	
	\author[First]{Marco Omainska} 
	\author[First]{Junya Yamauchi}
	\author[First]{Masayuki Fujita}
	
	\address[First]{Department of Information Physics \& Computing,
		The University of Tokyo, Tokyo, Japan (e-mail: marcoomainska@g.ecc.u-tokyo.ac.jp).}
	
	\begin{abstract} 
		This paper considers a scenario of pursuing a moving target that may switch behaviors due to external factors in a dynamic environment by motion estimation using visual sensors.
		First, we present an improved Visual Motion Observer with switched Gaussian Process models for an extended class of target motion profiles.
		We then propose a pursuit control law with an online method to estimate the switching behavior of the target by the GP model uncertainty.
		Next, we prove ultimate boundedness of the control and estimation errors for the switch in target behavior with high probability.
		Finally, a Digital Twin simulation demonstrates the effectiveness of the proposed switching estimation and control law to prove applicability to real world scenarios.
	\end{abstract}
	
	
\end{frontmatter}


\section{Introduction}
In robot control and Machine Learning visual sensors are a common tool for recognition tasks since they provide rich information \citep{Quintero:2017,Whalstrom:2015}.\\
Traditionally they are used for robot manipulators in vision-based control tasks, but recently shift to mobile (aerial) robots due to a wider range of modern applications \citep{spong_robot_2005}:
security and surveillance \citep{Pierson:2016tm},
investigation of animal ecology \citep{Shah2020},
bird strike prevention \citep{Paranjape2018},
among others.

\begin{figure}
	\centering
	\includegraphics[width=1\linewidth]{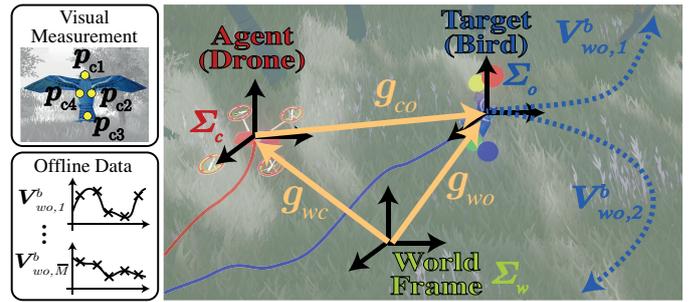}
	\caption{Pursuit of a \textit{target} (bird) by a camera equipped \textit{mobile robot} (drone) based on visual measurements and motion learning. The target can change its behavior (motion $\bodyvel[wo]$) anytime due to external influences.}
	\label{fig:digitwin}
\end{figure}

The scenario in this paper is the pursuit of a moving target based on visual measurements (\Cref{fig:digitwin}), which is a typical problem in the aforementioned applications.
Observers to estimate the target pose have been proposed in \cite{Quintero:2017,Chwa2016}, but
this paper focuses on the passivity-based Visual Motion Observer \citep{hatanaka_passivity-based_2015} since it simultaneously estimates the rigid body motion while pursuing the target.
To avoid the loss of vision to the target, previous research of the authors \cite{Omainska2021,Yamauchi2021} studied the inclusion of Gaussian Processes (GP), which are Bayesian non-parametric models with many favorable properties:
they provide an estimated mean value and measure their own model fidelity in form of a variance function \citep{Rasmussen:2006vz}.
Furthermore, upper bounds for the error between real and inferred function values based on the distance to training data do exist \citep{Srinivas:2012ez}.

In our previous work \citep{Omainska2021,Yamauchi2021} we assumed that targets move according to their environment.
A disadvantage about the proposed solutions is that the GP model can express only one particular motion trajectory.
This limits real world applicability, since other moving objects in a dynamic environment may influence the target behavior.
In contrast, \cite{Wei2019} consider multiple trajectories while targets leave and enter a scene.
A Dirichlet Process Gaussian Process estimates the probability of a target moving by a certain velocity field, but target motion switching which would require controller switching in our setting is not considered.
As such, \cite{Umlauft2020} online update the controller, but switching system dynamics are not considered.
Similarly, \cite{Maiworm2021} use a model predictive controller with recursive posterior prediction model updates, but switching system dynamics have not been considered.

In this paper, we consider the case of a target switching its motion in a dynamic environment.
We propose a method to online estimate the switching based on the GP model uncertainty and accordingly update the controller.
The contributions of this paper are as follows:
(i) extending the class of motion trajectories to target motion profiles,
(ii) proposing a visual pursuit control input with switched GP models and proving stability,
(iii) estimating the switching based on the GP variance,
(iv) showing applicability to real world scenarios with a Digital Twin simulation.

The paper is organized as follows:
First, the overall problem is formulated in \Cref{sec:problem}.
Next, \Cref{sec:vfs} introduces the Visual Pursuit System.
\Cref{sec:onlineswitchingcontroller} proposes the controller with online switching estimation method, which is then proven to be stable in \Cref{sec:stability}.
The effectiveness of the proposed controller is proven in a Digital Twin simulation in \Cref{sec:simulation}.
Finally, the conclusion is given in \Cref{sec:conclusion}.

\subsection*{Notation}
Vectors/matrices are denoted as bold characters.
$\norm{\cdot}$ is the Euclidean norm, $\eye[n]$ the identity matrix of size $n$ and $\succ$ shows positive definiteness of a matrix.
$\axang[ij]$ is the shorthand notation for $\eulax[ij]\theta_{ij}$.
Given two vectors $\boldsymbol{a}, \boldsymbol{b} \in \Rspace[3]$, the wedge-operator $\wedge$ computes $\hat{\boldsymbol{a}}\boldsymbol{b} = \boldsymbol{a} \times \boldsymbol{b}$ while the vee-operator $\vee$ is its inverse-operation.
The operator $\diag*{\cdot}$ defines a diagonal matrix with the given elements.
$\bracc{\cdot}_i$ and $\bracc{\cdot}_{i,i}$ denote elements or columns of a given vector or matrix.

\section{Problem Formulation} \label{sec:problem}
\subsection{Rigid Body Motion} \label{sec:rbm}
In this paper we consider the motion of rigid bodies with pose $\g \in \SEthree$ in the special Euclidean group
$\SEthree \coloneqq \Rspace[3] \times \SOthree$.
It consists of the position $\p \in \Rspace[3]$ and orientation $\rot \!\in\! \SOthree$. 
The axis-angle pair $\axang \in \Rspace[3]$ combines the normalized Euler axis $\eulax$ (i.e. $\norm*{\eulax} = 1$) and rotation angle $\theta \in (-\pi,\, \pi]$.
There are multiple representations of the pose $\g$,
but we will focus on the homogeneous representation \citep{hatanaka_passivity-based_2015}
\begin{equation} \label{eq:g}
	\g = \begin{bmatrix}
		\rot & \p \\
		\boldsymbol{0} & 1
	\end{bmatrix}
\end{equation}
and the vector form
$\gcheck = \begin{bmatrix}
	\p^\intercal & \eulax^\intercal \theta
\end{bmatrix}^\intercal$ \citep{Omainska2021}.
We consider the scenario of a moving target and camera-equipped mobile robot.
Let $\frame[w]$, $\frame[c]$ and $\frame[o]$ be coordinate frames of the world, camera, and target, respectively.
The motion of frame $\frame[j]$ with respect to $\frame[i]$ is then denoted as
\begin{equation} \label{eq:relativepose}
	\g[ij] \coloneqq \g[wi]^{-1}\g[wj] \, .
\end{equation}
Hence, we formulate the rigid body motion
\begin{equation} \label{eq:rbm}
	\dot{\g}_{wo} = \g[wo]\bodyvelH[wo] \, , \qquad \dot{\g}_{wc} = \g[wc]\bodyvelH[wc]
\end{equation}
of the target and camera with the body velocity
$\bodyvelH \!\in \sethree \coloneqq \curl{\begin{bsmallmatrix}
		\angvelskew & \transvel \\ \boldsymbol{0} & 0
\end{bsmallmatrix} \,\middle\vert\, \angvelskew \in \Rspace[3\times3], \, \para*{\angvelskew}^\intercal + \angvelskew = \boldsymbol{0}, \, \transvel \in \Rspace[3]}$
where $\transvel$ corresponds to the translational and $\angvel$ to the angular velocity, respectively.
The relative rigid body motion (RRBM) from the camera to the target can be obtained from the time derivative of \cref{eq:relativepose} with \cref{eq:rbm} as
\begin{equation} \label{eq:RRBM}
	\dot{\g}_{co} = -\bodyvelH[wc]\g[co] + \g[co]\bodyvelH[wo] \, .
\end{equation}

\subsection{Switched Target Motion Profiles} \label{sec:targetmotion}
The target moves according to the unknown body velocity
$\bodyvel[wo] = (\transvel[wo], \, \angvel[wo]) \in \Rspace[6]$
with the following assumption:
\begin{assumption} \label{ass:motion}
	The target moves in a bounded field $D \!\subset\! \Rspace[3]$, thus, $\Xset \coloneqq D \times \SOthree$ is a compact subset of $\SEthree$.
\end{assumption}
Note that targets naturally move in bounded environments, hence \Cref{ass:motion} is non-restrictive.
Further, targets are moving under external influences, i.e. terrain and other moving obstacles, and may switch behavior at times.

Previous research \citep{Omainska2021,Yamauchi2021} assumed that targets move under the influence of a static environment at which $\bodyvel[wo]$ results in only one periodic trajectory.
However, this limits the applicability to real scenarios since it does not take into account dynamic environments that may force the target to switch its behavior.
Hence, similar to the definition of a position-based velocity field in \cite{Wei2019} we introduce
\textit{target motion profiles}
$\bodyvel[wo,\psi] \colon \Xset \rightarrow \Rspace[6]$
drawn from a broader set
\begin{equation} \label{eq:bodyvelclass}
	\bodyvelclass \coloneqq \curl*{\bodyvel[wo,1],\,\ldots\,,\bodyvel[wo,\bar{M}]}
\end{equation}
where index $\psi \in \Psi \coloneqq \curl{1 \, , \, \ldots \, , \, \bar{M}}$ is determined at time $t \in [0,\, \infty)$ by an unknown switching signal \citep{Liberzon2003}
\begin{equation} \label{eq:s(t)}
	s(t) \colon \left[0, \, \infty\right) \rightarrow \Psi
\end{equation}
resulting in a piece-wise continuous body velocity signal
$\bodyvel[wo](t,\gcheck[wo]) = \bodyvel[wo,s(t)](\gcheck[wo])$.
Note that we exclude infinite switching in finite time cases to prevent Zeno behavior.

Our goal is to follow a moving target \cref{eq:rbm} under switched motion trajectories \cref{eq:bodyvelclass}.
It is assumed that the camera knows its pose $\g[wc]$, however, since $\g[co]$ cannot be directly measured, $\g[wo]$ cannot be calculated from \cref{eq:relativepose} either.
A new control law is proposed in \Cref{sec:onlineswitchingcontroller} that improves pursuit under the new target motion profiles \cref{eq:bodyvelclass}.
We also prove boundedness even if the switching signal \cref{eq:s(t)} is unknown and show a method how it can be estimated online.

\section{Preliminaries: Visual Pursuit System} \label{sec:vfs}
\subsection{Visual Measurements} \label{sec:visualmesurements}
In the following we will discuss how the mobile robot gathers information of a moving target through its visual sensor.
We assume targets have at least $n_f \geq 4$ feature points (i.e. yellow markers in \Cref{fig:digitwin}) which can be extracted by real time Computer Vision techniques such as given in \cite{Kane2020}.
The positions $\p[oi] \in \Rspace[3],\, i \in \curl{1, \,\ldots\, , n_f}$ of the detected feature points in target frame $\frame[o]$ can be transformed into camera frame $\frame[c]$ by a coordinate transformation $\begin{bmatrix}\p[ci]^\intercal & 1\end{bmatrix}^\intercal = \g[co]\begin{bmatrix}\p[oi]^\intercal & 1\end{bmatrix}^\intercal$.
The positions $\p[ci] = \begin{bmatrix}x_{ci} & y_{ci} & z_{ci}\end{bmatrix}^\intercal$ are then projected onto the image plane as
$\vismea[i] = \tfrac{\lambda}{y_{ci}} \begin{bmatrix} x_{ci} & z_{ci} \end{bmatrix}^\intercal \in \Rspace[2]$
with $\lambda > 0$ the focal length of the camera \citep{spong_robot_2005}.
By stacking all $\vismea[i]$ we obtain the \textit{visual measurement} \cite[p.105]{hatanaka_passivity-based_2015}
\begin{equation} \label{eq:visual_measurement}
	\vismea(\g[co]) = \begin{bmatrix}\vismea[1]^\intercal & \ldots & \vismea[{n_f}]^\intercal\end{bmatrix} \in \Rspace[2n_f] \ .
\end{equation}

\subsection{Visual Pursuit Control} \label{sec:vmo}
\begin{figure}
	\centering
	\includegraphics[width=1\linewidth]{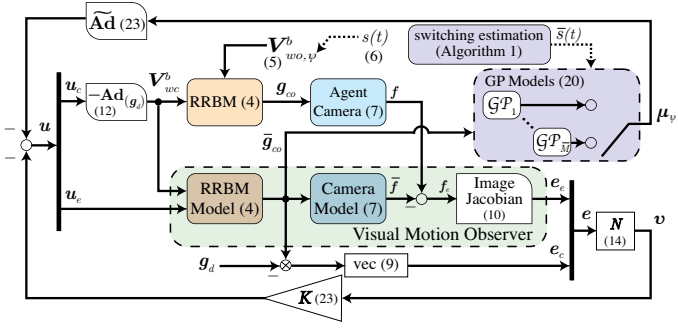}
	\caption{Block diagram of the visual pursuit system with VMO \cref{eq:vmo}, proposed control law \cref{eq:control_law}, $\GPmodel[\psi]$ models \cref{eq:GPmodel} and switching estimation~\Cref{alg:switching_estimation}.}
	\label{fig:vfs}
\end{figure}
In order to follow the target we have to simultaneously measure its pose $\g[wo] = \g[wc]\g[co]$ from visual information.
Since $\g[co]$ is unavailable, we define an estimate $\gbar[co]$ of the relative rigid body motion \cref{eq:RRBM} given by the \textit{Visual Motion Observer} (VMO)
\cite[Sec.6]{hatanaka_passivity-based_2015}
\begin{equation} \label{eq:vmo}
	\dot{\gbar}_{co} = -\bodyvelH[wc]\gbar[co] - \gbar[co]\hat{\u}_e
\end{equation}
with the observer input $\u[e]$.
We are interested in driving $\g[co]$ to a desired pose $\g[d] \in \SEthree$ while rendering $\gbar[co]$ closer to $\g[co]$.
Thus, let us define the control and estimation errors
\begin{equation} \label{eq:control_estimation_error}
	\begin{aligned}
		\g[ce] &\coloneqq \g[d]^{-1}\gbar[co], \quad \e[c] \coloneqq \vectrafo{\g[ce]}\\
		\g[ee] &\coloneqq \gbar[co]^{-1}\g[co], \quad \e[e] \coloneqq \vectrafo{\g[ee]}
	\end{aligned}
\end{equation}
with $\vectrafo{\g} \!=\! \begin{bmatrix}\p^\intercal & {\skew*{\rot}^\vee}^\intercal\end{bmatrix}^\intercal$ and
$\skew*{\rot} \!=\! (\rot - \rot[-])/2$.
But we cannot compute $\g[ee]$ due to its dependency on $\g[co]$.
Therefore, based on \cref{eq:visual_measurement} we will compute an estimated visual measurement $\bar{\vismea}$ from $\gbar[co]$ and define the visual measurement error $\vismea[e] = \vismea - \vismea[][\bar]$.
Let us make the following assumption:
\begin{assumption} \label{ass:bounded_angle}
	Both the control and estimated rotation error are bounded as $\abs{\theta_{ce}(t)} < \pi/2,\, \abs{\theta_{ee}(t)} < \pi/2,\, \forall t \geq 0$.
\end{assumption}
This allows us to calculate the estimation error $\g[ee]$ from
\begin{equation}
	\e[e] = \jacobian^\dagger(\gbar[co])\vismea[e]
\end{equation}
with $\jacobian^\dagger$ the pseudo-inverse of the \textit{image Jacobian} $\jacobian$ \citep[p.108]{hatanaka_passivity-based_2015}.
Also note that the assumption on $\theta_{ce}$ is in our scenario generally satisfied since the mobile robot must be able to move faster than the target for successful pursuit.
Now, the time derivative of \cref{eq:control_estimation_error} yields
\begin{equation}
	\begin{aligned}
		\dot{\g}_{ce} &= \hat{\u}_c\g[ce] - \g[ce]\hat{\u}_e \\
		\dot{\g}_{ee} &= \hat{\u}_e\g[ee] + \g[ee]\bodyvelH[wo]
	\end{aligned}
\end{equation}
with $\u[c] = -\adjoint[\g[d]^{-1}]\bodyvel[wc]$.
Then, with the \textit{adjoint matrix}
\begin{equation}
	\adjoint[\g] = \begin{bmatrix}
			\rot & \hat{\p}\rot \\
			\boldsymbol{0} & \rot
		\end{bmatrix} \ ,
\end{equation}
control and estimation error systems $\bodyvel[ce] \coloneqq (\g[ce]^{-1}\dot{\g}_{ce})^\vee$
and $\bodyvel[ee] \coloneqq\ (\g[ee]^{-1}\dot{\g}_{ee})^\vee$,
we obtain the overall \textit{error system}
\begin{equation} \label{eq:error_system}
	\begin{bmatrix}
		\bodyvel[ce] \\ \bodyvel[ee]
	\end{bmatrix} =
	\begin{bmatrix}
		\adjoint[\g[ce]^{-1}] & -\eye[6] \\
		\boldsymbol{0} & \adjoint[\g[ee]^{-1}]
	\end{bmatrix}
	\u +
	\begin{bmatrix}
		\boldsymbol{0} \\ \eye[6]
	\end{bmatrix}
	\bodyvel[wo] \ .
\end{equation}
The input $\u \coloneqq \begin{bmatrix}\u[c]^\intercal & \u[e]^\intercal\end{bmatrix}^\intercal \in \Rspace[12]$ is to be designed later in \Cref{sec:onlineswitchingcontroller}.
Let us also define the stacked error vector $\e \coloneqq \begin{bmatrix}\e[c]^\intercal & \e[e]^\intercal\end{bmatrix}^\intercal \in \Rspace[12]$ and output of the error system \cref{eq:error_system}
\begin{equation} \label{eq:output}
	\nuu \coloneqq \N\e, \quad \N \coloneqq \begin{bmatrix}
		\eye[6] & \boldsymbol{0} \\
		-\adjoint[{\rot[-ce]}] & \eye[6]
	\end{bmatrix} \ .
\end{equation}
The \textit{visual pursuit system} is illustrated in \Cref{fig:vfs}.

\section{Online Switching Control Law} \label{sec:onlineswitchingcontroller}
\subsection{Gaussian Process Model} \label{sec:gp}
Consider noisy measurements of \cref{eq:bodyvelclass} in the form of
\begin{equation} \label{eq:datameasurement}
	\y = \bodyvel[wo,\psi](\x) + \eps_\psi, \quad \x \in \Xset,\ \y \in \Rspace[6]
\end{equation}
with sub-Gaussian noise $\eps_\psi \sim \Nspace*{0}{\boldsymbol{\sigma}_{n,\psi}^2}$ and standard deviation $\boldsymbol{\sigma}_{n,\psi} \!\coloneqq\! \diag*{\sigma_{n,\psi,1},\,\ldots\,,\,\sigma_{n,\psi,6}}$.
Measurements are stored in $\bar{M}$ datasets with individually $M_\psi$ datapoints
\begin{equation} \label{eq:dataset}
	\Dset_\psi \!=\! \curl{\!\Xdata_\psi, \Ydata_\psi\!},
	\begin{array}{l}
		\begin{aligned}
			\Xdata_\psi &\!\coloneqq\! \bracc{\x^{\curl*{1}}\, \cdots\, \x^{\curl*{\! M_\psi \!}}}^\intercal \!\!\in\! \Rspace[M_\psi \times 6] \\
			\Ydata_\psi &\!\coloneqq\! \bracc{\y^{\curl*{1}}\, \cdots\, \y^{\curl*{\! M_\psi \!}}}^\intercal \!\!\in\! \Rspace[M_\psi \times 6]
		\end{aligned}
	\end{array}
\end{equation}

On a compact set the \textit{squared exponential (SE)} kernel
\begin{equation} \label{eq:kernel}
	\kernel_\psi(\x, \x^\prime) = \sigma_{f,\psi}^2 \exp{-\tfrac{1}{2}(\x-\x^\prime)^\intercal\LambdaGP_\psi(\x-\x^\prime)}
\end{equation}
computes the correlation between two inputs $(\x, \x^\prime)$ and approximates any continuous function arbitrarily precisely \citep{Seeger2008}.
It measures properties of \cref{eq:datameasurement} together with the symmetric lengthscales matrix $\LambdaGP_\psi$ and signal variance $\sigma_{f,\psi}$ which are typically obtained by evidence maximization \citep{Rasmussen:2006vz}.
Finally, the prediction $\y^*$ at test input $\x^*$ is jointly Gaussian distributed with prior mean zero and the posterior mean and variance for every output $i=1, \ldots, 6$ are defined as
\begin{align}
	\mu_{\psi,i}(\x^*) &= \extendedcov_\psi^\intercal \para*{\Gram_\psi +\boldsymbol{\sigma}_{n,\psi}^2\eye[M_\psi]}^{-1}\bracc*{\Ydata_\psi}_i \label{eq:meanGP} \\
	\sigma^2_{\psi,i}(\x^*) &= \kernel_\psi(\x^*,\x^*) - \extendedcov_\psi^\intercal \para*{\Gram_\psi + \boldsymbol{\sigma}_{n,\psi}^2\eye[M_\psi]}^{-1}\extendedcov_\psi \label{eq:covGP}
\end{align}
The Gram matrix is computed as $\bracc*{\Gram_\psi}_{j,j^\prime} \!=\! \kernel_\psi(\Xdata_{\psi}^{\curl*{j}}\!,\Xdata_{\psi}^{\curl*{j^\prime}})$ for $j,j^\prime \in \curl*{1,\, \ldots,\, M_\psi}$ and further
$\bracc*{\extendedcov_{\psi}}_j  = \kernel_\psi(\Xdata_{\psi}^{\curl*{j}},\x^*)$.
The combined multi-variable Gaussian distribution is then
\begin{equation} \label{eq:GPmodel}
	\begin{aligned}
		\meanGP_\psi &= \bracc{\mu_{\psi,1} \ \cdots \ \mu_{\psi,6}}^\intercal \in \Rspace[6] \\
		\varGP_\psi &= \diag{\sigma^2_{\psi,1}, \, \ldots \, , \sigma^2_{\psi,6}} \in \Rspace[6\times6]
	\end{aligned}
\end{equation}
which we will call the $\GPmodel[\psi]$ model for motion profile $\bodyvel[wo,\psi]$.
\begin{remark}
	The data \cref{eq:dataset} is usually hard to obtain online due to real time computation constraints and noisy measurements.
	However, \cite{Omainska2021} has proven that we can use the VMO to observe the data.
	Dividing the dataset into individual target motion profiles has the benefit that conflicts in the data (i.e. crossing points in the target trajectory) can be avoided that would else result in wrong predictions of \cref{eq:meanGP} and higher noise estimates \cref{eq:datameasurement}.
\end{remark}

\subsection{Control Law with Switched Motion Trajectories} \label{sec:controller}
Consider the common Lyapunov function candidate
\begin{equation} \label{eq:storage}
	S_\psi \coloneqq \tfrac{1}{2} \sum_{j \in \curl{c,e}} \para{\norm*{\p[je]}^2 + \trace*{\eye[3]-\rot[je]}} 
\end{equation}
with the time derivative \citep{hatanaka_passivity-based_2015}
\begin{equation} \label{eq:storage_derivative}
	\dot{S}_\psi = \nuu^\intercal \u + \e^\intercal \begin{bmatrix} \boldsymbol{0} \\ \adjoint[\rot[ee]] \end{bmatrix} \bodyvel[wo,\psi] \ .
\end{equation}
When the target does not move ($\bodyvel[wo,\psi] \!\equiv\! 0$) the system \cref{eq:error_system} is passive from input $\u$ to output $\nuu$ with respect to the storage function \cref{eq:storage}.
Thus, under \Cref{ass:bounded_angle} and switched $\GPmodel[\psi]$ models \cref{eq:GPmodel} we propose the control law
\begin{equation} \label{eq:control_law}
	\u = -\K\nuu - \adjoint[][\widetilde]\meanGP_{\bar{\psi}}
\end{equation}
with $\adjoint[][\widetilde] \coloneqq \begin{bmatrix}\adjoint[\rot[ce]]^\intercal & \eye[6]\end{bmatrix}^\intercal \adjoint[\rot[ee]]$
and controller gains $\K = \diag{\K_c, \, \K_e}, \ \ \K_c, \!\K_e \in \Rspace[6\times6]$.
The choice of \GPmodel[\psi] in \cref{eq:control_law} depends on the online estimate $\bar{\psi} = \bar{s}(t), \, \forall t \in [0, \, \infty)$ of the real switching function \cref{eq:s(t)} and will be discussed next.

\subsection{Switching Estimation}
In the author's previous work \citep{Omainska2021,Yamauchi2021} a method to adapt controller and communication gains based on the GP uncertainty has been studied.
This motivates us in this paper to propose \Cref{alg:switching_estimation} to online estimate the switching $\bar{\psi} = \bar{s}(t), \, \forall t \in [0, \, \infty)$
based on the minimum uncertainty
\begin{equation} \label{eq:switching_estimation}
	\bar{\psi} = \argmin_{\psi \in \Psi} \frac{\norm*{\alphaswitch_{\psi}^\intercal\varGP_{\psi}^{\sfrac{1}{2}}}}{\bar{\Sigma}_{\alphaswitch_{\psi}}}
\end{equation}
with the normalization factor $\bar{\Sigma}_{\alphaswitch_{\psi}} = \max_{\x \in \Xset} \norm*{\alphaswitch_{\psi}^\intercal\varGP_\psi^{\sfrac{1}{2}}(\x)}$ since the $\GPmodel[\psi]$ models do in general not share equal hyper-parameters and noise $\eps_\psi$.
The design parameter $\alphaswitch_{\psi} \in \Rspace[6]$ specifies the importance of some output dimensions in $\varGP_{\psi}$ on the switching detection.
Further, to circumvent abundant switching between $\GPmodel[\psi]$ models of approximately equal variance, the constant parameter $T \in [0,\, 1)$ is introduced in \Cref{alg:switching_estimation}.
At $T=0$ switching happens immediately, while for higher $T$ switching happens only if the difference in variance is large enough.
\begin{algorithm2e}[t]
	\caption{Switching Estimation of \cref{eq:s(t)}} \label{alg:switching_estimation}
	\SetKwInput{init}{Initialize}
	
	\KwOut{switching estimate $\bar{\psi}^* = \bar{s}(t)$}
	\KwIn{$\varGP_\psi$, $\alphaswitch_{\psi}$, $\bar{\Sigma}_{\alphaswitch_{\psi}}$ $\forall \psi \in \Psi$ and $T \in [0,\,1)$}
	\init{ $\bar{\psi}^*$ from \cref{eq:switching_estimation}}
\BlankLine
\While{controller \cref{eq:control_law} running}
{
	obtain $\bar{\psi}$ from \cref{eq:switching_estimation}\;
	\If{$\sfrac{\norm{\alphaswitch_{\bar{\psi}^*}^\intercal\varGP_{\bar{\psi}^*}^{\sfrac{1}{2}}}}{\bar{\Sigma}_{\alphaswitch_{\bar{\psi}^*}}} > \sfrac{\norm{\alphaswitch_{\bar{\psi}}^\intercal\varGP_{\bar{\psi}}^{\sfrac{1}{2}}}}{\bar{\Sigma}_{\alphaswitch_{\bar{\psi}}}} + T$}
	{
		$\bar{\psi}^* \gets \bar{\psi}$\;
	}
}
\end{algorithm2e}

\section{Stability Analysis} \label{sec:stability}
In this section we will derive the main theorems of this paper.
First, a bound of always using the least accurate $\GPmodel[\psi]$ model (worst-case) for unknown switching \cref{eq:s(t)} will be derived.
Then, an individual bound for every $\GPmodel[\psi]$ model is derived when \cref{eq:s(t)} can be inferred well.

\subsection{Ultimate Boundedness for Unknown Switching}
Let us assume the following:
\begin{assumption} \label{ass:Vbwo_bounded}
	The target motion profiles \cref{eq:bodyvelclass} are bounded, i.e. $\norm*{\bodyvel[wo,\psi](\x)} < \infty, \ \forall \x \in \Xset, \ \forall \psi \in \Psi$.
\end{assumption}
We are now ready to state the first theorem about the worst-case scenario when the switching \cref{eq:s(t)} is unknown:
\begin{theorem} \label{thm:switching_unknown}
	Suppose that \Cref{ass:bounded_angle,ass:motion,ass:Vbwo_bounded} hold.
	Consider the proposed control law \cref{eq:control_law} with $\bar{M}$ trained \GPmodel[\psi] models \cref{eq:GPmodel} of $M_\psi$ datapoints \cref{eq:dataset} each.
	Then, the error $\norm{\e}$ of the error system \cref{eq:error_system} is uniformly ultimately bounded with regards to arbitrary switching signals \cref{eq:s(t)} and estimate \cref{eq:switching_estimation}.
\end{theorem}
\begin{proof}
	Starting from \cref{eq:storage_derivative} by inserting the control law \cref{eq:control_law} 
	\begin{equation} \label{eq:thm1_proof_eq1}
		\begin{aligned}[b]
			\dot{S}_\psi
			&= -\nuu^\intercal \K\nuu + \e^\intercal
					\begin{bmatrix} \boldsymbol{0} \\ \adjoint[{\rot[ee]}] \end{bmatrix} (\bodyvel[wo,\psi] - \meanGP_{\bar{\psi}}) \\
			&\leq -\nuu^\intercal \K\nuu + \norm*{\e[e]} \norm*{\bodyvel[wo,\psi] - \meanGP_{\bar{\psi}}}
		\end{aligned}
	\end{equation}
	and using Cauchy-Schwarz inequality in the last equation.
	Since the kernel \cref{eq:kernel} is continuous, mean \cref{eq:meanGP} is bounded.
	Together with \Cref{ass:Vbwo_bounded} it follows that we can find an\\ upper bound $\maxmodelerror \!\coloneqq\!\! \displaystyle\max_{\psi,{\bar{\psi}} \in \Psi}\norm*{\bodyvel[wo,\psi] \!-\! \meanGP_{\bar{\psi}}}$ such that with \cref{eq:output}
	\begin{equation}
		\begin{aligned}[b]
			\dot{S}_\psi
			&\leq -\e^\intercal \N^\intercal \K \N\e + \norm*{\e[e]} \maxmodelerror \\
			&\leq -\lambda_{\K} \norm*{\e}^2 + \norm*{\e[e]} \maxmodelerror
		\end{aligned}
	\end{equation}
	where $\lambda_{\K} > 0$ is the smallest eigenvalue of $\N^\intercal\K\N$.
	Finally, from $\norm*{\e}^2 = \norm*{\e[c]}^2 + \norm*{\e[e]}^2$ and quadratic extension
	\begin{equation}
		\begin{aligned}
			\dot{S}_\psi
			\leq -\lambda_{\K}\norm*{\e[c]}^2 - \lambda_{\K}\para*{\norm*{\e[e]} - \tfrac{1}{2\lambda_{\K}} \maxmodelerror}^2 + \tfrac{1}{4\lambda_{\K}} \maxmodelerror^2
		\end{aligned}
	\end{equation}
	it is shown that the storage function \cref{eq:storage} decreases $\dot{S}_\psi < 0$ $\forall \g[wo] \in \Xset \backslash \bar{\Bset}, \ \forall \psi \in \Psi$ outside the set
	\begin{equation} \label{eq:ellipse_max}
		\bar{\Bset} \coloneqq \curl*{\forall \g[wo] \in \Xset \, \mid \, \sqrt{\norm*{\e[c]}^2 + \para*{\norm*{\e[e]} - c}^2} \leq c}
	\end{equation}
	which forms an ellipse in $\e$ with constant $c \!=\!\! \frac{1}{2\lambda_{\K}}\maxmodelerror$.
	From \cite[Theorem 2.1]{Liberzon2003}
	and the fact that $S_\psi$ is a common Lyapunov function $\forall \psi \!\in\! \Psi$,
	it follows that the error is uniformly ultimately bounded with regards to arbitrary switching signals \cref{eq:s(t)} and converges to ellipse \cref{eq:ellipse_max}.
\end{proof}

Note that \Cref{thm:switching_unknown} assures stability for unknown $s(t)$ only with a worst-case bound.
We will discuss next a more fitting result when $s(t)$ is known.

\subsection{Bound for Known Switching}
In order to quantify the \GPmodel[\psi] model error $\modelerror$ we need to have some prior knowledge about the target motion profiles \cref{eq:bodyvelclass}.
Hence, we extend \Cref{ass:Vbwo_bounded} to the following:
\begin{assumption} \label{ass:RKHS}
	Every target motion profile of \cref{eq:bodyvelclass} has a bounded reproducing kernel Hilbert space (RKHS) norm $\norm*{\bracc*{\bodyvel[wo,\psi]}_i}_{\kernel_\psi}$ associated with kernel \cref{eq:kernel} with known hyperparameters for every output $i = 1,\ldots,6$.
\end{assumption}
This assumption is difficult to verify, however, it only excludes discontinuities and other irregularities within the class \cref{eq:bodyvelclass}.
We note that even if the overall target body velocity can include discontinuities, \Cref{ass:RKHS} can be still satisfied by the proposed splitting into different target motion profiles $\bodyvel[wo,\psi]$ that are continues by themselves.
\begin{lemma} \label{lm:lipschitz}
	Suppose \Cref{ass:RKHS} holds. Then there exists $L_\psi \!>\! 0$ such that
	$\abs*{\bracc*{\bodyvel[wo,\psi](\x)}_i-\bracc*{\bodyvel[wo,\psi](\x^\prime)}_i} \!\leq\! L_\psi \norm{\x \!-\! \x^\prime}$.
\end{lemma}
\begin{proof}
	For the sake of this proof we will use shortened notations $V_i \!=\! \bracc*{\bodyvel[wo,\psi]}_i$, $\kernel_{\x} \!=\! \kernel(\x,\cdot)$, and $\norm*{\x}_{\LambdaGP} \!=\! \sqrt{\x^\intercal\LambdaGP_\psi\x}$.
	From the reproducing property
	$\dotprod*{V_i}{\kernel_{\x}}_{\kernel} \!=\! V_i(\x), \, \forall \x \!\in\! \Xset$ we get
	$\abs*{V_i(\x) - V_i(\x^\prime)}^2 \!=\! \abs*{\dotprod{V_i}{\kernel_{\x} \!-\! \kernel_{\x^\prime}}_{\kernel}}^2 \!\leq\! \norm*{V_i}_{\kernel}^2 d_{\kernel}^2(\x,\x^\prime)$\\
	where Cauchy-Schwarz is used for the inequality.
	We aim to show that the distance over the metric $\kernel$ is upper bounded by
	$d_{\kernel}^2(\x,\x^\prime) = \kernel_{\x}(\x) - 2\kernel_{\x^\prime}(\x) + \kernel_{\x^\prime}(\x^\prime) \leq \sigma_{f,\psi}^2\norm*{\x-\x^\prime}^2_{\LambdaGP}$
	which is easily proven by substituting $r\coloneqq\norm*{\x-\x^\prime}_{\LambdaGP}$ and \cref{eq:kernel}. We finally obtain
	$\abs*{V_i(\x) - V_i(\x^\prime)}^2 \leq L_\psi^2\norm*{\x-\x^\prime}^2$
	with a Lipschitz constant
	$L_\psi^2 = \sigma_{f,\psi}^2\lambda_{max}(\LambdaGP_\psi)\norm*{V_i}_{\kernel}^2$.
\end{proof}
Thus, we use \cite[Lemma 4.2]{Omainska2021} to obtain
\begin{equation} \label{eq:Lipschitz}
	\norm*{\bodyvel[wo,\psi](\gcheck[wo]) \!-\! \bodyvel[wo,\psi](\gcheckbar[wo])} \!\leq\! \Lip[p,\psi] \norm{\p[ee]} \!+\! 2\pi\Lip[\theta,\psi]
\end{equation}
with Lipschitz constants $\Lip[p,\psi]$, $\Lip[\theta,\psi]$.
According to \cite{Srinivas:2012ez} we can also state the following:
\begin{lemma} \label{lm:boundedProbability}
	Let \Cref{ass:RKHS} hold, then each model error
	\begin{equation} \label{eq:GPerror}
		\modelerror(\x) = \norm*{\bodyvel[wo,\psi](\x) - \meanGP_\psi(\x)}, \quad \forall \psi \in \Psi
	\end{equation}
	is bounded by a probability $\delta \in (0,\,1)$ as
	\begin{equation} \label{eq:boundedProbability}
		\Probspace*{\modelerror(\x) \leq \norm*{\betaGP_\psi^\intercal \varGP_\psi^{\sfrac{1}{2}}(\x)}, \ \forall \x \in \Xset} \geq 1-\delta^6
	\end{equation}
	on the compact set $\Xset$ where each $\betaGP_\psi \in \Rspace[6]$ is computed as
	$\bracc*{\betaGP_\psi}_i \!=\! \sqrt{2\norm*{\bracc*{\bodyvel[wo,\psi]}_i}_{\kernel_\psi}^2 + 300\bracc*{\maxinfoGP_\psi}_i \log^3(\sfrac{(M_\psi+1)}{\delta})}$,
	based on the maximum information gain $\maxinfoGP_\psi \in \Rspace[6]$ for $i = 1,\ldots,6$.
\end{lemma}
\begin{remark}
	It comes natural to assume $\alphaswitch_{\psi} \equiv \betaGP_\psi$ as a good candidate in \Cref{alg:switching_estimation} since $\betaGP_\psi$ details the amount of information in all outputs of the $\GPmodel[\psi]$ model. In fact, \cref{eq:switching_estimation} can then be interpreted as the minimization over the bound \cref{eq:boundedProbability} for $\GPmodel[\psi]$ model error \cref{eq:GPerror} by a certain probability $1-\delta^6$.
\end{remark}
We are now ready to state the main theorem.
\begin{theorem} \label{thm:switching_known}
	Suppose that \Cref{ass:bounded_angle,ass:motion,ass:RKHS} hold and the switching \cref{eq:s(t)} with $\bar{\psi}=\psi$ is known.
	Consider the proposed control law \cref{eq:control_law} with $\bar{M}$ trained \GPmodel[\psi] models~\cref{eq:GPmodel} of $M_\psi$ datapoints \cref{eq:dataset} each.
	Suppose $\tilde{\lambda}_{\K} \coloneqq \lambda_{\K}-\Lip[p,\psi] > 0$
	with $\lambda_{\K}$ the minimum eigenvalue of $\N^\intercal\K\N$ and the Lip-schitz constants $\Lip[p,\psi]$, $\Lip[\theta,\psi]$ from \cref{eq:Lipschitz}.
	Then, the error $\norm*{\e}$ of the error system \cref{eq:error_system} is uniformly ultimately bounded and converges by a probability $\delta \in (0,\, 1)$ to an ellipse
	\begin{align} \label{eq:ellipse_psi}
			\Bset_\psi &\coloneqq \curl{\forall \g[wo] \in \Xset \, \middle| \, E \leq 0} \\
			E &\coloneqq \sqrt{\norm{\e[c]}^2 + \tfrac{\tilde{\lambda}_{\K}}{\lambda_{\K}}\para{\norm{\e[e]} - c(\varGP_\psi)}^2}
			- \sqrt{\tfrac{\tilde{\lambda}_{\K}}{\lambda_{\K}}}c(\varGP_\psi) \notag
	\end{align}
	with
	$c(\varGP_\psi) = \tfrac{1}{2\tilde{\lambda}_{\K}}\norm*{\betaGP_\psi^\intercal \varGP_\psi^{\sfrac{1}{2}}}+\tfrac{\pi\Lip[\theta,\psi]}{\tilde{\lambda}_{\K}}$.
\end{theorem}
\begin{proof}
	Since the real $\g[wo]$ is not available, we have to use its estimation for the $\GPmodel[\psi]$ mean prediction $\meanGP_\psi(\gcheckbar[wo])$:
	\begin{equation} \label{eq:thm2_proof_eq1}
		\begin{aligned}
			\dot{S}_\psi
			\begin{multlined}[t]
				\leq -\nuu^\intercal \K\nuu + \norm{\e[e]} \left(\norm*{\bodyvel[wo,\psi](\gcheckbar[wo]) - \meanGP_\psi(\gcheckbar[wo])}\right. \\
				\left. + \norm*{\bodyvel[wo,\psi](\gcheck[wo]) - \bodyvel[wo,\psi](\gcheckbar[wo])}\right)
			\end{multlined}
		\end{aligned}
	\end{equation}
	We used the triangle inequality in \cref{eq:thm2_proof_eq1} and from $\tilde{\lambda}_{\K} > 0$, $\norm{\p[ee]} \leq \norm{\e[e]}$, \cref{eq:Lipschitz,eq:GPerror} the following inequality holds for a probability $\delta \in (0,\, 1)$ and constant factor $d=2\pi\Lip[\theta,\psi]$:
	\begin{equation}
		\dot{S}_\psi \leq -\lambda_{\K}\norm*{\e[c]}^2 -\tilde{\lambda}_{\K}\para{\norm*{\e[e]}-c(\varGP_\psi)}^2 + \tilde{\lambda}_{\K}c(\varGP_\psi)^2
	\end{equation}
	It follows that the storage function \cref{eq:storage} decreases $\dot{S}_\psi < 0$ $\forall \g[wo] \in \Xset \backslash \Bset_\psi, \ \forall \psi \in \Psi$ outside set \cref{eq:ellipse_psi} for probability $\delta$.\\
	Since $c(\varGP_\psi)$ is upper-bounded, uniform ultimate boundedness follows analogous to \Cref{thm:switching_unknown}.
\end{proof}
\begin{remark}
	\Cref{thm:switching_unknown} proves the existence of a bound for error $\e$ for all times, however, the maximum error $\maxmodelerror$ is in\linebreak[4]
	practice difficult to calculate.
	On the contrary, \Cref{thm:switching_known} focuses on providing specific bounds for every $\psi \in \Psi$ individually and by a certain probability.
	We note that even if $\bar{\psi} = \bar{s}(t)$ is poorly estimated, after an initial convergence period $\e$ is bounded between \cref{eq:ellipse_max} and \cref{eq:ellipse_psi}.
	Further, we note for $T>0$ in \Cref{alg:switching_estimation} Zeno behavior is excluded.
\end{remark}

The constant factor in \cref{eq:Lipschitz} originates from a worst case bound since in general $\axang[wo] - \bar{\eulax}\bar{\theta}_{wo} \neq \axang[ee]$.
However, in many scenarios rotations are restricted to only one dimension, i.e. the axis of rotation is often fixed as the vertical axis.
Let us thus make the following assumption:
\begin{assumption} \label{ass:rotaxis}
	The rotation axis $\eulax[wo]$ is known and fixed.
\end{assumption}
With \Cref{ass:rotaxis} we can fix all initial orientations to $\eulax \equiv \bracc*{0 \ 0 \ 1}^\intercal$.
This simplifies \cref{eq:Lipschitz} with a Lipschitz constant
$\norm*{\bodyvel[wo,\psi](\gcheck[wo]) - \bodyvel[wo,\psi](\gcheckbar[wo])} \leq \Lip[\psi] \norm{\e[e]}$
and we obtain:
\begin{corollary} \label{cor:axis_known}
	Let \Cref{ass:rotaxis} and the same conditions as in \Cref{thm:switching_known} hold.
	Then, for $\tilde{\lambda}_{\K} \coloneqq \lambda_{\K} - \Lip[\psi] > 0$, the ellipse \cref{eq:ellipse_psi} is simplified with $c(\varGP_\psi) = \frac{1}{2\tilde{\lambda}_{\K}}\norm*{\betaGP_\psi^\intercal \varGP_\psi^{\sfrac{1}{2}}}$.
\end{corollary}
\begin{proof}
	Follows similar to \Cref{thm:switching_known} with the new $\Lip[\psi]$.
\end{proof}

With \Cref{cor:axis_known} the region where $\dot{S}_\psi < 0$ entirely depends on the $\GPmodel[\psi]$ model quality.
That means, the size of the ellipse \cref{eq:ellipse_psi} depends only on the \GPmodel[\psi] variance $\varGP_\psi$ and shrinks with more data and a smaller noise variance.


\section{Simulation} \label{sec:simulation}
In this section we will illustrate how the proposed control law \cref{eq:control_law} performs in a Digital Twin simulation\footnotemark{}.

\subsection{Setup}
\begin{figure}
	\centering
	\includegraphics[trim=0 0 0 0.2cm,clip,width=1\linewidth]{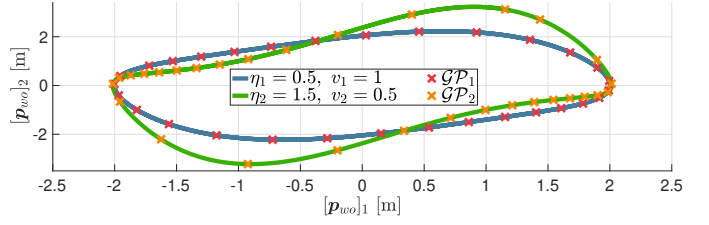}
	\vspace{-1.8em}
	\caption{Bird trajectories resulting from motion profiles \cref{eq:simulation_targetmotion}. The crosses denote the data for two $\GPmodel[\psi]$ models \cref{eq:GPmodel}.}
	\label{fig:trajectory}
\end{figure}
\footnotetext[\value{footnote}]{The code is available at:\\ \url{https://github.com/marciska/vpc-switched-motion}}
The given scenario (\Cref{fig:digitwin}) of a target ("bird") and a pursing mobile robot ("drone") is implemented in Unity, whilst the control input and motion estimation is calculated in \MATLAB{}.
Unity and \MATLAB{} communicate over a ROS interface in a Docker-composed network with a message frequency of $\SI{50}{\Hz}$.
For a practical physics simulation, we use the rigid body component of Unity.

The target moves according to the velocity ($\psi=1,2$)
\begin{equation} \label{eq:simulation_targetmotion}
	\begin{aligned}
		\transvel[wo,\psi](\gcheck[wo]) &= \rot[-wo] \begin{bmatrix} a_x(\p[wo]) & a_y(\p[wo]) & 0 \end{bmatrix}^\intercal \\
		\angvel[wo,\psi](\gcheck[wo]) &= \begin{bmatrix} 0 & 0 & \dv{t}\atantwo(a_x(\p[wo]), a_y(\p[wo])) \end{bmatrix}^\intercal \\
		a_x(\p[wo]) &= v_\psi\bracc{\p[wo]}_2 \\
		a_y(\p[wo]) &= -v_\psi\bracc{\p[wo]}_1 + v_\psi\eta_\psi(1-\bracc{\p[wo]}{}_1^2)\bracc{\p[wo]}_2
	\end{aligned}
\end{equation}
with $\eta_1 = 0.5$, $v_1 = 1$ and $\eta_2 = 1.5$, $v_2 = 0.5$ resulting in two Van der Pol oscillators as shown in \Cref{fig:trajectory} with the target orientation always heading towards the front path.
We learn $\bar{M}=2$ $\GPmodel[\psi]$ models with $M_\psi=30$ data points each, where the data is collected according to \cref{eq:datameasurement} with noise variance $\sigma^2_{n,i} = 0.01^2$, $i=1,\ldots,6$.
Since the target always moves in line of sight we set $\alphaswitch_{\psi} = \bracc*{0 \ 1 \ 0 \ 0 \ 0 \ 0}^\intercal$ and $T=0.05$ in \Cref{alg:switching_estimation} as the bird forward motion includes the most information for switching detection.
The simulation is run for $\SI{20}{\second}$ which results in 2 rounds around the trajectory center, while the switching happens at $\p[wo] = \bracc{\pm2 \ 0 \ 0}$.

The Lipschitz constant $\Lip[\psi] \leq \bar{\Lip}$ is approximated by an upper bound $\bar{\Lip} = 8$ for both trajectories.
Then, with the controller gains $\K_c = 10\eye[6]$, $\K_e = 17\eye[6]$ and since the target orientation is limited to rotations around the vertical axis, the condition $\tilde{\lambda}_{\K} > 0$ from \Cref{cor:axis_known} is satisfied with $\lambda_{\K} = 10$.
The rigid body poses are initialized at
$\gbar[co](0) \!=\! (\begin{bmatrix} 0 & 1 & 0 \end{bmatrix}^\intercal,\,\eye[3])$,
$\g[wo](0) \!=\! (\begin{bmatrix} -2 & 0 & 0 \end{bmatrix}^\intercal,\,\eye[3])$,
$\g[wc](0) \!=\! (\begin{bmatrix} -2 & -3 & 0 \end{bmatrix}^\intercal,\,\eye[3])$,
and $\g[d] \!=\! (\begin{bmatrix} 0 & 2 & 0 \end{bmatrix}^\intercal,\,\eye[3])$.
We are interested in the below candidates for the proposed control law \cref{eq:control_law}:
\begin{enumerate}[left=1em, label=\textit{Case \arabic*}:, ref=\textit{Case \arabic*}]
	\item\label{case:fullGP} Single $\GPmodel$ model trained on full dataset ($M_1+M_2=60$ data points, no $\GPmodel$ switching) from \cite{Omainska2021}
	\item\label{case:switchedGP} Switched \GPmodel[\psi] with estimation \Cref{alg:switching_estimation}
\end{enumerate}
\vspace{-0.13em}

\subsection{Results}
\begin{figure}
	\centering
	\includegraphics[trim=0 4.5cm 5cm 4.5cm,clip,width=1\linewidth]{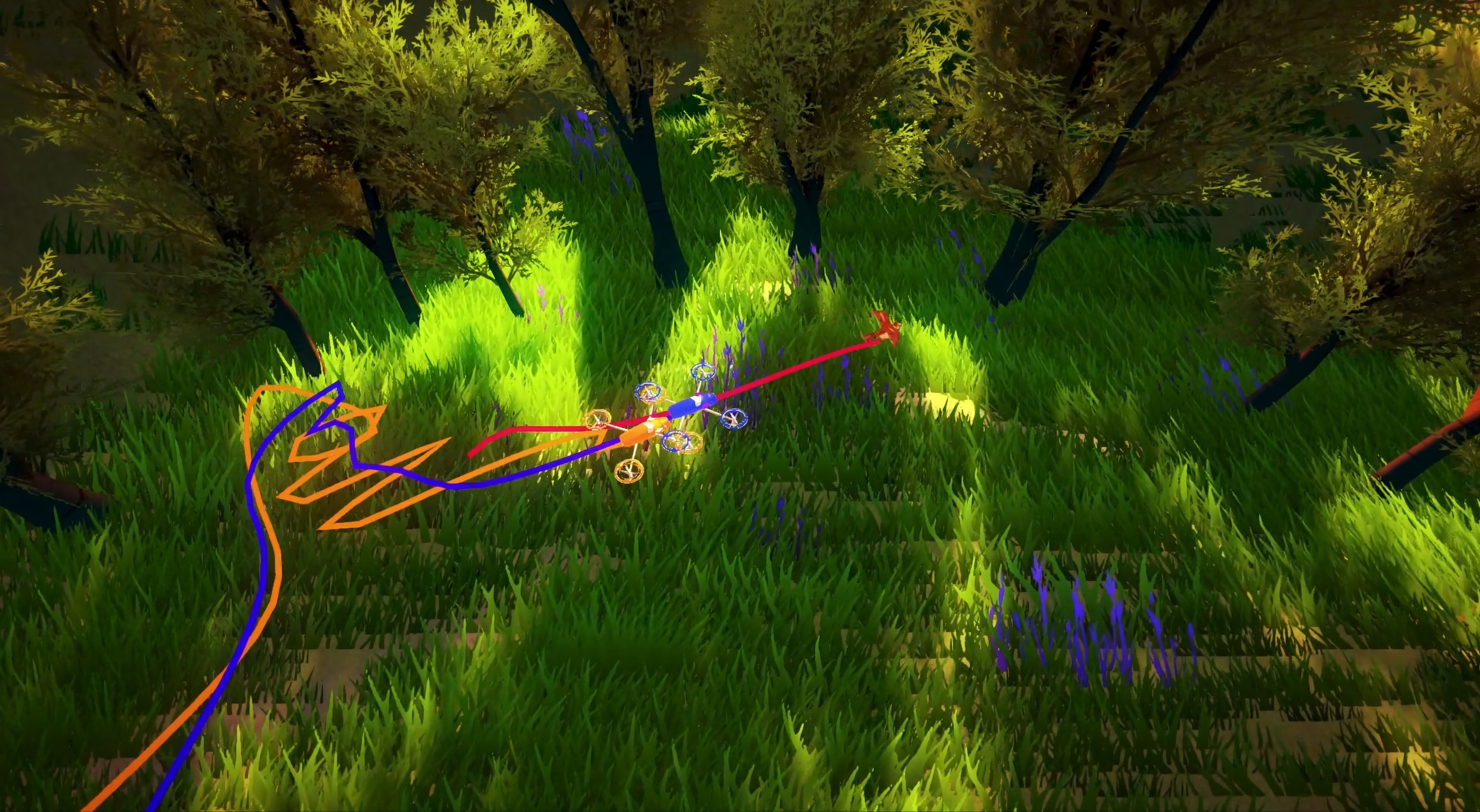}
	\vspace{-1.2em}
	\caption[Snapshot of simulation]{%
		Simulation snapshot: \small{\url{https://youtu.be/YxX8FoeyF8g}}%
	}
	\label{fig:snapshot_simulation}
\end{figure}
\begin{figure}
	\centering
	\includegraphics[trim=0.8cm 0.4cm 2.9cm 1.1cm,clip,width=1\linewidth]{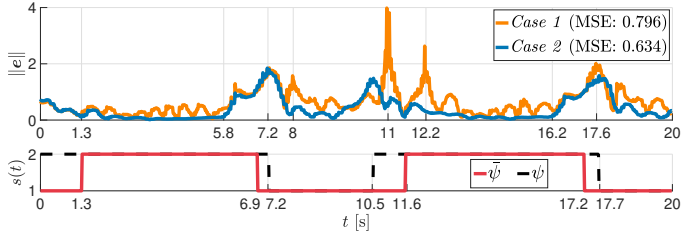}
		\vspace{-1.5em}
	\caption{%
		Top: Comparison of error $\norm*{\e}$.
		Bottom: Estimated switching $\bar{s}(t)$ (solid) and real $s(t)$ (dashed).%
	}
	\label{fig:results}
\end{figure}
\Cref{fig:results,fig:snapshot_simulation} show the simulation results.
It is clear that \Cref{case:switchedGP} significantly improves the pursuit performance.
As such, \Cref{case:fullGP} shows an overall oscillatory behavior (orange trail in \Cref{fig:snapshot_simulation}) with high error peaks at $t=\SI{11}{\second}$ and $t=\SI{12.2}{\second}$ in \Cref{fig:results} as a result of $\GPmodel$ misprediction.
These are especially prevalent at points where the trajectories overlap, since the velocity \Cref{eq:simulation_targetmotion} on the trajectories \Cref{fig:trajectory} is different and thus the $\GPmodel$ model has to overcome a mismatch in the data.
This hypothesis is further strengthened by the fact that the \GPmodel\ hyperparameters obtained from evidence maximization are significantly larger in the respective outputs than for \Cref{case:switchedGP}, and hence the $\GPmodel$ seeks to predict both velocities by a certain trade-off.
\Cref{case:switchedGP} on the other hand reveals a good performance even when the trajectories overlap, since the distinct motion profiles \Cref{eq:simulation_targetmotion} have been separated into different $\GPmodel[\psi]$ models and misprediction can only happen in the case of a bad switching estimate $\bar{s}(t)$.\linebreak[4]
Nonetheless, despite the initial misprediction at simulation start, the results show that \Cref{alg:switching_estimation} can sufficiently estimate the switching function \cref{eq:s(t)}.
Moreover, the oscillatory behavior of \Cref{case:fullGP} is not preserved in \Cref{case:switchedGP}.

Overall, \Cref{case:switchedGP} (Mean Square Error / MSE: $0.634$) shows a $20.35\%$ performance improvement when compared to \Cref{case:fullGP} (MSE: $0.796$).
In summary, \Cref{case:switchedGP} demonstrates that it is important to include target switching behavior in dynamic environments in the control law \cref{eq:control_law} and that the switching can be well estimated by the proposed \Cref{alg:switching_estimation}.

\section{Conclusion} \label{sec:conclusion}
This paper proposes a visual pursuit control law based on a motion observer and switched Gaussian Process models.
A new class of target motion profiles is derived that the target can exhibit at times influenced by a dynamic environment.
We propose a method to estimate the motion switching based on the GP variance.
For every target motion profile a unique error bound is derived.
Finally, we show that the proposed control law effectively estimates the switching motion and pursues a target in a Digital Twin simulation.


\bibliography{root}

\end{document}